\documentclass[journal]{IEEEtran}
\IEEEoverridecommandlockouts
\usepackage{cite}
\usepackage{amsmath,amssymb,amsfonts}
\usepackage{algorithmic}
\usepackage{graphicx}
\usepackage{textcomp}
\usepackage{xcolor}
\usepackage{hyperref}
\def\BibTeX{{\rm B\kern-.05em{\sc i\kern-.025em b}\kern-.08em
    T\kern-.1667em\lower.7ex\hbox{E}\kern-.125emX}}
\begin{document}



    \title{When Geoscience Meets Generative AI and Large Language Models: Foundations, Trends, and Future Challenges}
  \author{Abdenour Hadid, 
  Tanujit Chakraborty, and~Daniel Busby
  
\thanks{ A. Hadid is with the Sorbonne Centre for Artificial Intelligence, Sorbonne University Abu Dhabi, UAE. (e-mail: abdenour.hadid@sorbonne.ae).}
\thanks{Corresponding Author: T. Chakraborty is with the Department of Science and Engineering, Sorbonne University Abu Dhabi, UAE, and also with the Sorbonne Centre for Artificial Intelligence, Paris, France. (e-mail: tanujit.chakraborty@sorbonne.ae).}
\thanks{D. Busby is with TotalEnergies, PAU, France. (e-mail: daniel.busby@totalenergies.com).}
\thanks{A. Hadid and T. Chakraborty have equal contribution.}}  


\maketitle


\begin{abstract}
Generative Artificial Intelligence (GAI) represents an emerging field that promises the creation of synthetic data and outputs in different modalities. GAI has recently shown impressive results across a large spectrum of applications ranging from biology, medicine, education, legislation, computer science, and finance. As one strives for enhanced safety, efficiency, and sustainability, generative AI indeed emerges as a key differentiator and promises a paradigm shift in the field. This paper explores the potential applications of generative AI and large language models in geoscience. The recent developments in the field of machine learning and deep learning have enabled the generative model's utility for tackling diverse prediction problems, simulation, and multi-criteria decision-making challenges related to geoscience and Earth system dynamics. This survey discusses several GAI models that have been used in geoscience comprising generative adversarial networks (GANs), physics-informed neural networks (PINNs), and generative pre-trained transformer (GPT)-based structures. These tools have helped the geoscience community in several applications, including (but not limited to) data generation/augmentation, super-resolution, panchromatic sharpening, haze removal, restoration, and land surface changing. Some challenges still remain such as ensuring physical interpretation, nefarious use cases, and trustworthiness. Beyond that, GAI models show promises to the geoscience community, especially with the support to climate change, urban science, atmospheric science, marine science, and planetary science through their extraordinary ability to data-driven modeling and uncertainty quantification.
\end{abstract}


\section{Introduction} \label{into}
\begin{figure}[t!] 
\centering
\includegraphics[width=0.48\textwidth]{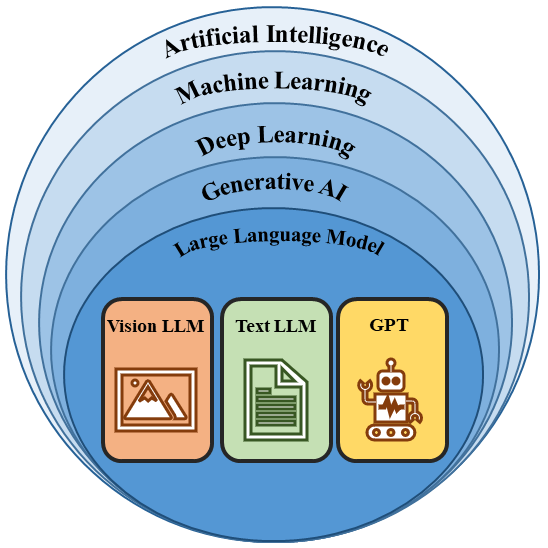}
\caption{The relation between large language models (LLMs), generative AI, and other related learning schemes.}
\label{relations}
\end{figure}

Among the most fascinating topics that have recently attracted the attention of researchers, enthusiasts, and even the general public is the {\it Generative Artificial Intelligence} (GAI). The foremost ability of GAI lies in creating new content such as text, images, and audio, and hence fostering creativity. The main purpose of such models is to generate new samples from what was already there in the training data. This new technological development has pushed the boundaries of what machines can achieve. As illustrated in Fig. \ref{relations}, generative AI is a subset of deep learning methods that can work with both labeled and unlabeled data of varied modalities using supervised, unsupervised, or semi-supervised learning schemes. However, the first idea of generative models dates back to the 1950s with the  introduction of Gaussian mixture models (GMMs)~\cite{cao2023comprehensive} and hidden Markov models (HMMs)~\cite{eddy1996hidden}. One of the earliest applications of these tools was in the field of speech recognition~\cite{rabiner1989tutorial}. After the rise of deep learning, the productivity and capability of generative models have enhanced significantly. Deep learning models are a class of AI models that can be either {\it discriminative} or {\it generative}. Discriminative models are typically trained on a large dataset of labeled samples, aiming to discover the relationship between the features of the samples and the given labels. Once a discriminative model is trained, it can be utilized to predict the label of any new sample. A generative model, on the other hand, creates new sample instances based on a learned probability distribution of existing samples. 

To create new content using prompt engineering, GAI models are built using large-scale datasets during training. This boosts GAI models to produce excellent results by recognizing patterns in the data based on a probability distribution and then creating ``similar" patterns when prompted. GAI can run on different models that use different mechanisms to do training and output new content. These include, for example, generative adversarial networks (GANs)~\cite{goodfellow2014generative}, diffusion models~\cite{sohl2015deep, yang2023diffusion}, transformers~\cite{vaswani2017attention}, variational autoencoders (VAEs)~\cite{kingma2013auto} and Large Language Models (LLMs)~\cite{radford2019better, zhao2023survey}. One of the major breakthroughs in GAI is the introduction of GANs in 2014~\cite{goodfellow2014generative} in which one network generates content (generative model) and the other helps in figuring out whether it is an authentic sample or not (discriminative model). The next breakthrough came with the idea of Transformers~\cite{vaswani2017attention} in the field of natural language processing (NLP) and computer vision (CV), followed by the creation of BERT~\cite{devlin2018bert} and GPT~~\cite{radford2018improving}. Most recently, the ChatGPT model was introduced, and the first version was created by OpenAI~\cite{ray2023chatgpt}. These GAI models have shown impressive capabilities across a broad spectrum of tasks, much beyond showcasing its generative power in writing emails and reports. It has, in fact, attracted increasing attention and has been recently deployed in numerous downstream applications, encompassing diverse fields such as biology and medicine~\cite{singhal2023large,singhal2023towards,lievin2022can,nori2023capabilities, sharma2023performance, antaki2023evaluating, levine2023diagnostic}, education~\cite{tack2022ai,wang2023chatgpt,pardos2023learning,dai2023can}, legislation~\cite{savelka2023explaining,chalkidis2020legal,blair2023can,yu2022legal}, computer programming~\cite{ross2023programmer,sandoval2023lost,leinonen2023comparing,xu2022systematic,thapa2022transformer,liu2023your}, and finance~\cite{fieberg2023using,zaremba2023chatgpt,son2023beyond,wu2023bloomberggpt,araci2019finbert,zhang2023xuanyuan}. New applications of generative AI are appearing every day.

The rapid progress in artificial intelligence has been a game-changer across various industries, and geoscience is no exception. The integration of GAI in geoscience is revolutionizing the way geoscientists interpret and understand the Earth’s complex, interactive, and multiscale processes~\cite{zhang2023geoscience}. This fusion of technology and science is proving to be a catalyst for innovation and progress, offering unprecedented opportunities for breakthroughs in the field of geoscience. The availability of massive volumes of geo and Earth system data is encouraging the applications of machine learning and deep learning tools for data-driven geoscience, Earth science, and remote sensing~\cite{patel2023generative}. In such scenarios, AI is proven to handle the complexity and ambiguity of geophysical data better than classical geostatistical methods. Geophysical data is often noisy and incomplete, making it challenging to interpret. AI algorithms, particularly deep learning models, are designed to handle such complexities. They can identify patterns and relationships in the data that humans might misread, leading to more accurate and reliable interpretations. For instance, AI can predict seismic activities by analyzing patterns in historical data, thereby helping in disaster management~\cite{karpatne2018machine}. Furthermore, AI can assist in resource exploration by identifying optimal sites for oil and gas extraction~\cite{bergen2019machine}. This not only increases efficiency but also reduces the environmental impact of these activities. Perhaps one of the most recent and notable examples of generative AI is ChatGPT - a language model developed by OpenAI (\url{https://openai.com/blog/chatgpt}). It is trained on a diverse range of texts and is capable of generating coherent and contextually relevant content. While ChatGPT-like models are commonly used for tasks like generating text, answering questions, and providing recommendations, their application in geophysical engineering is a relatively unexplored topic with great potential. On the other hand, the integration of AI in geoscience is not without challenges. One of the main hurdles is the lack of unbiased, high-quality, labeled data for training AI models. Geophysical data is often unstructured and heterogeneous, making it difficult to use for machine learning algorithms, and limiting generalization.

Despite the rapid development, GAI is still in its early stages, as the success of these models is clearly tied to the quantity of the training data associated with the foundation model. Also, from the trustworthiness and fairness viewpoints, some serious concerns have been raised about the potential emergence of ``super-intelligent machines" without adequate safeguards. Generative AI can indeed unintentionally produce biased or incorrect information due to biases in its training data. Moreover, generative models usually require a large amount of high-quality, unbiased data to operate. Other equally important issues include the latency for generating high-quality samples, the massive computing power that is needed to train generative models, and the maxim of ``garbage in and garbage out''. All these issues are important to be taken into consideration to effectively capitalize and guarantee the beneficial use of generative AI in the field of geoscience and beyond. This survey discusses the potential of generative AI in geoscience, providing some guidance, highlighting the challenges, and pointing out some future research directions.

The rest of this article is structured as follows. Section \ref{applications} discusses and enumerates the potential of generative AI and large language models in geoscience. Then, an overview of existing GAI models and tools in geoscience are thoroughly presented in Section \ref{tools}. Section \ref{benchmarks} describes some useful benchmark datasets and resources for geoscience GAI. Discussion and conclusions are drawn in sections \ref{discussion} and \ref{conclusion}, respectively.


\section{Geoscience in the Era of Generative AI} \label{applications}
\begin{figure}[t!] 
\centering
\includegraphics[width=0.50\textwidth]{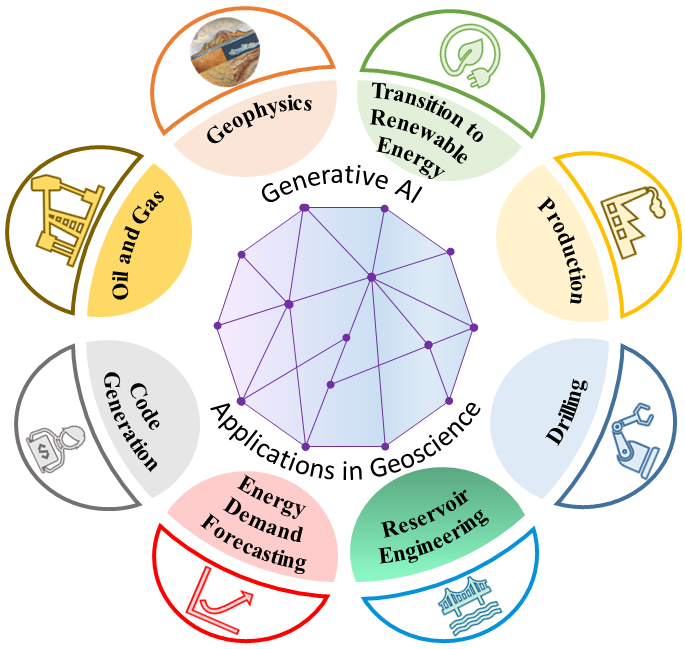}
\caption{Some potential application domains of generative AI in geoscience. }
\label{Fig_applications}
\end{figure}

\noindent This section discusses some potential applications of generative AI in geoscience. The most natural use is perhaps through generating “new” samples for training the AI-based models. This is useful in applications like reservoir modeling in which geoscientists are often required to deal with several samples of the subsurface and develop probabilistic models to help with the subsequent decision-making process. Examples of current attempts at using generative AI in geoscience include but are not limited to, the reconstruction of 3D porous media~\cite{mosser2017reconstruction}, the generation of geologically realistic 3D reservoir facies using deep learning of sedimentary architecture~\cite{zhang2019generating}, and seismic data augmentation~\cite{wang2021seismogen} using seismic waveform synthesis and GANs. More details on seismic data augmentation modeling ("SeismoGen") are given in Section~\ref{GAN_Geoscience} as a use case of using generative AI in geoscience. Another use case of large language models (called {\it K2}) for understanding and answering geoscience questions is discussed in Section~\ref{Foundation_Geoscience}. The {\it K2} model is claimed to be able to generate new geoscience ideas as well. As illustrated in Fig.~\ref{Fig_applications}, the diverse application of GAI in geoscience provides unique opportunities for characterizing the Earth systems. Data collected from different resources plays a critical role in building GAI systems. Several non-exhaustive potentials of generative AI in geoscience include:\\

\noindent {\bf Reservoir Engineering}: Generative AI can be used in reservoir engineering to examine fluid flow, well performance, and production data.\\ 
{\bf Reservoir Characterization and Modeling}: LLMs could assist in data interpretation. 
By analyzing vast amounts of geological and geophysical data, the models can offer valuable insights to help engineers make decisions.\\
{\bf Production Optimization:} LLMs can assist in providing real-time analysis of production data and suggesting adjustments to extraction rates, pressure levels, and other operational parameters.\\ 
{\bf Drilling}: LLMs could be used to analyze geological data and predict drilling risks, which are important for good planning. Moreover, the models can also help determine optimal drilling parameters and provide insights into drilling fluid properties and their effects on drilling efficiency. Additionally, GPT-based models can help monitor drilling operations, identify possible problems, and suggest solutions.\\
{\bf Production}: LLMs can analyze production data and identify potential production problems aiming for optimal performances.\\ 
{\bf Geomechanics}: By analyzing geophysical data, GPT-based models can provide useful information about rock strength and permeability, for instance. It can also help predict potential geomechanical risks concerning subsidence, fault reactivation, and induced seismicity.\\
{\bf Code Generation}: It is well known that GAI models are good at code generation as they can automatically generate programming scripts from high-level descriptions given by a user (``programmers''). This is partially possible thanks to the advanced natural language processing capabilities of LLM models. However, the automatically produced code may not be always fully understandable.\\ 
{\bf Carbon Emissions Reduction and Transition to Renewable Energies:} As recently discussed at the 2023 United Nations Climate Change Conference (COP28) in the United Arab Emirates, generative AI can indeed help fulfill the compliance standards for carbon emissions by developing cleaner production processes~\cite{romanello2023further}. 

\section{Overview of Generative AI Models in Geoscience} \label{tools}

\begin{figure*}[htb] 
\centering
\includegraphics[width=0.98\textwidth]{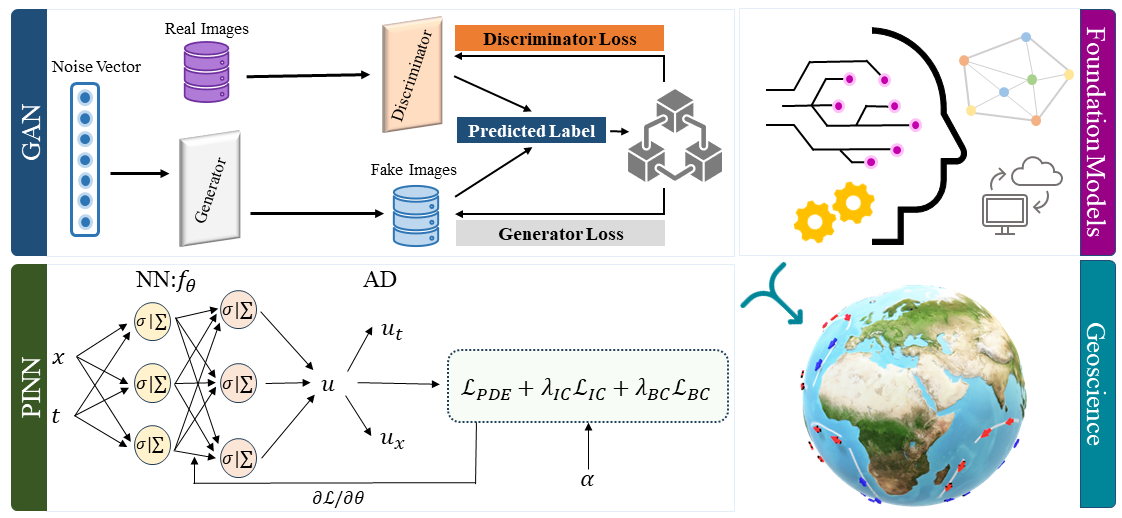}
\caption{Popularly-used Generative AI models for geoscience applications. These deep learning (DL) frameworks learn features automatically from the data by extracting information from it. GANs work as an unsupervised framework, whereas PINNs are hybrid physics+DL models for modeling system dynamics and solving differential equations. Foundation models mainly focus on in-context learning by using a pre-trained architecture. These models are widely applicable to geoscience applications.}
\label{GAI_Models_Geosciece}
\end{figure*}

In the past, machine learning and deep learning models have been widely used for solving geo and Earth science problems~\cite{artificialhandbook}. GAN-based GAI models and LLM-based tools have found interesting applications in the field of geoscience for various functionalities. To incorporate geoscience domain knowledge into AI systems, physics-informed neural networks (PINNs)-based methodologies combine robust domain knowledge of the dynamics along with deep learning frameworks for modeling and simulation. A pictorial view of these GAI methods for geoscience is portrayed in Fig. \ref{GAI_Models_Geosciece}. 


\subsection{GAN-based Architectures}\label{GAN_Geoscience}
GANs~\cite{goodfellow2014generative} emerged as a fundamentally novel approach for solving synthetic data generation problems in computer vision applications. GANs framework establishes an adversarial min-max game between its generator and discriminator networks. During training, the generator module endeavors to produce realistic synthetic data by sampling from a probability distribution, such as Gaussian, and utilizes a neural network as a universal approximator function to map the data domain. Conversely, the discriminator module is trained to differentiate synthetic data from real data samples, creating an adversarial training dynamic between the networks. In the realm of geoscience, GANs have become indispensable for diverse applications, including generating realistic seismic images~\cite{wang2021seismogen}, simulating geological structures~\cite{zhang2019generating}, augmenting remote sensing data~\cite{lv2021remote}, designing road networks in futuristic cities~\cite{thottolil2023prediction}, and reconstructing 3-dimensional porous media~\cite{mosser2017reconstruction}, among others. This section discusses several GAN-based models and their applicability in geoscience contexts.

\textbf{SeismoGen for Earth Data.}
Introduced by Wang {\it et al.}~\cite{wang2021seismogen}, the SeismoGen framework seeks to enhance the generalization capabilities of earthquake detection methods based on machine learning and deep learning by supplementing real Earth data with synthetic seismic waveforms. This GAN-based model demonstrates the capacity to generate three-component waveforms with multiple labels. Utilizing the conditional GAN framework~\cite{mirza2014conditional}, the SeismoGen architecture enables the incorporation of domain-specific knowledge into synthetic data. The generator in the SeismoGen framework is composed of a four-layered convolutional neural network, designed to generate synthetic waveform data and corresponding labels using a Gaussian noise vector and input labels. The discriminator network, on the other hand, consists of two sequential modules, namely the feature extractor and the sample critic. The former module of the network learns a feature representation from input seismic signals and the latter one provides critiques based on the learned features. To optimize the minimax game between the generator and discriminator, and facilitate a stable learning process, the value function of the Wasserstein GAN~\cite{arjovsky2017wasserstein} is employed. Thus, the generator loss function ($L_G$) and the discriminator loss function ($L_D$) can be mathematically expressed as:
\begin{align*}
L_G=&-\mathbb{E}_{z \sim \operatorname{N}(0,1)} D(G(z)),\\
L_D=&\mathbb{E}_{z \sim \operatorname{N}(0,1)} D(G(z))-\mathbb{E}_{x \sim p_{\text{R}}} D(x) \\ &+\lambda \mathbb{E}_{z \sim \operatorname{N}(0,1)}\left[\left(\|D(G(z))\|_2-1\right)^2\right],
\end{align*}
where $z$ represents the Gaussian noise vector, $x$ represents the real data from $p_{\text{R}}$, and $\lambda$ is the hyperparameter. The primary goal is to minimize the disparity between authentic seismic waveforms and those synthetically generated by SeismoGen. This involves iterative optimization of $L_G$ and $L_D$ to establish equilibrium between the generator and discriminator networks. The conditional generation feature of SeismoGen has proven its capability to produce highly realistic labeled seismic waveforms, proving valuable for waveform analysis and data augmentation~\cite{li2021residual}. 

\textbf{GAI Tools for Urban Science.}
Application of GANs in urban science became a reality with the help of CityGAN~\cite{Albert2018ModelingUP}, which learns architectural features of major cities and produces images of futuristic buildings that do not exist. CityGAN is used to generate a ``urban universe'' from global urban land-use imageries. Another variant of CityGAN, Metropolitan GAN (MetroGAN)~\cite{zhang2022metrogan} is based on progressively growing~\cite{karras2017progressive} least square loss function~\cite{mao2017least} and incorporates geographical loss and enforces constraints based on physical geography. The robust urban morphology simulation capabilities of the MetroGAN covering several commendable strengths, however, encounter difficulties in efficiently representing all intricate features of complex urban systems. RidgeGAN combines a kernel ridge regression (KRR)~\cite{vovk2013kernel, murphy2012machine} with the generative CityGAN to address the problem of imaging future cities of India~\cite{thottolil2023prediction}. RidgeGAN captured the urban morphology of selected Indian small and medium-sized cities and derived its transportation network index using human settlement indices. This hybrid GAI model is built on open-source datasets, namely WSF2019 and OSM data~\cite{marconcini2020outlining}. In the hybridization, KRR was used to predict the transportation index of Indian cities of the future. This urban science-based GAI approach appeared valuable in situations where real data for developing and underdevelopment notions is limited, as it can generate data closely resembling the actual urban morphology. However, there exist several limitations of these tools for urban science applications, such as validation of the simulated data and dealing with time-varying urban patterns as human settlement changes over time.

\textbf{GANs for Geological Facies Modeling.}
Geological facies modeling is an essential foundation for exploring and characterizing sub-surface reservoirs. Traditional geostatistical models have been widely used within the geoscientific community for decades to generate sub-surface models. Recent advancements in the field of GAI enabled the use of GANs combined with training-image-based simulation for geological facies modeling~\cite{zhang2019generating}. Zhang {\it et al.}extended GANs to 3D geological modeling at the reservoir scale to generate a wide variety of geologically realistic facies models constrained by well data interpretations. Their geomodeling approach has been validated on models such as complex fluvial depositional systems and carbonate reservoirs. This framework uses both perceptual loss and conceptual loss in the loss function where contextual loss is calculated using a distance transformation which measures the dissimilarity between GAN-generated sample and facies observation at each well location. To incorporate uncertainty quantification within GANs, a Bayesian GAN was introduced to create a facies model with an increasing computational cost~\cite{feng2022application}. Bayesian GANs have been applied to several geological depositional scenarios to capture the variability of the data. U-Net GAN framework emerged as an alternative solution for subsurface geological facies with fragmentary measurements~\cite{zhang2021u}. This study used deep convolutional GANs to produce unconditional facies models and the U-Net to model conditional geological facies. However, most of these methods do not consider the scales of the geological features, which remains a future challenge for the geoscientific community~\cite{song2021geological}.

\subsection{Foundation Models in GeoScience}\label{Foundation_Geoscience}
\begin{figure*}[htb] 
\centering
\includegraphics[width=0.98\textwidth]{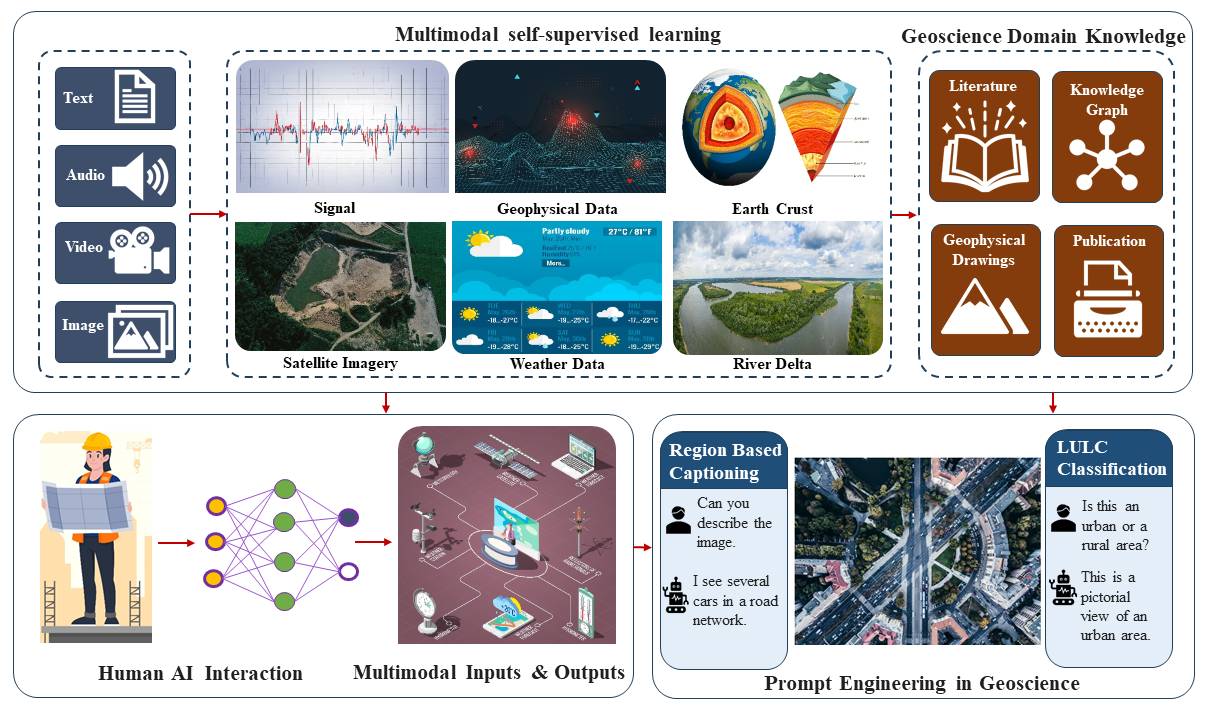}
\caption{Overview of a Generative AI model pipeline for Geoscience Applications. Multimodal self-supervised learning algorithms train various data types (image, text, speech, numerical data) using multidimensional geoscience data models from satellite imagery, weather, earth observations, and rivers. An example of prompt engineering using GeoChat is represented (right below) for demonstration.}
\label{GAI_Overview}
\end{figure*}

Foundation models in geoscience possess the capabilities to tackle challenges in multimodal and multidimensional datasets with numeric, text, audio, and video inputs. GAI models employ a scientific representation of geoscience domain knowledge that enables us to engage in reasoning processes and provides insights for geoscientists. The key benefits of using foundation models in geoscience are scalability (data size, model training, and computational power do not make a problem), generalizability (superior performance on new tasks going beyond the training data), dynamic interactions (convolutional operations and multimodality), and flexibility~\cite{zhang2023geoscience}. A schematic diagram representing the workflow of foundation models in geoscience is presented in Fig. \ref{GAI_Overview}.

\textbf{K2: An LLM Model for Geoscience.}
K2 is one of the first open-source LLMs specialized in geoscience~\cite{k2}. Large language models, such as ChatGPT, are indeed known to yield impressive results in general-purpose language understanding and generation. They acquire this amazing ability by learning from massive amounts of training data. However, general-purpose LLMs may not be good enough in specialized topics such as geoscience. To design a specialized LLM for geoscience, one needs to (1) collect a large amount of geoscience knowledge, (2) consider a general-purpose LLM and train it with the geoscience corpus, and 3) fine-tune the model with supervised geoscience data. This is basically the idea behind K2 proposed by Deng {\it et al.}~\cite{k2}. More specially, the authors in~\cite{k2} trained a general-purpose LLaMA model~\cite{touvron2023llama} with 5.5B tokens of geoscience literature (Wikipedia pages, geoscience paper’s abstracts, open-access geoscience papers from top journals in Earth science, and other sources from Internet), and utilized geoscience-supervised data (called GeoSignal) to fine-tune the model. Experiments on a benchmark dataset, called GeoBench, showed that the {\it K2} model is able to answer geoscience questions and is useful for research assistance and knowledge reasoning. {\it K2} is also claimed to be able to generate new geoscience ideas. The model was compared favorably against similar-size baseline models such as Galactica-6.7B ~\cite {taylor2022galactica}, MPT-7B~\cite {team2023introducing}, Vicuna-7B~\cite {chiang2023vicuna}, LLaMA-7B~\cite {touvron2023llama} and Alpaca-7B ~\cite{taori2023stanford}. Such domain-specific LLM is undoubtedly valuable for engineers and researchers in geoscience. 

\textbf{GeoChat: A Vision LLM.} This recent advancement embodies a revolutionary shift in geoscience through the integration of Large Vision Language Models (Vision LLM), reshaping the analysis and comprehension of Earth's intricate dynamics. Introduced by Kukreja {\it et al.}\cite{kuckreja2023geochat}, the GeoChat framework, a robust Vision LLM-based remote sensing (RS) architecture, exhibits exceptional performance across various RS tasks, including image and region captioning, visual interrogation, scene classification, referring detection, and visually grounded conversations. This architecture employs an innovative data generation pipeline, utilizing object detection datasets~\cite{wang2023samrs} for image description and Vicuna-V1.5~\cite{chiang2023vicuna} for text-based conversations. Additionally, visual question-answering and scene classification capabilities are integrated using respective RS datasets. Structural enhancements in GeoChat include a task token component specifying the desired task type, such as grounding, image level, or region-level conversation. The framework also incorporates spatial location representation for precise location specification. Visual representation, generated by pre-trained CLIP-ViT (L-14)~\cite{tay2017learning}, is projected onto the language model space using a single-layered neural network adaptor with GeLU activation function~\cite{hendrycks2016gaussian}. Language tokens are then processed in Vicuna-V1.5~\cite{chiang2023vicuna}, serving as a unified interface for diverse Vision LLM inputs. Efficient fine-tuning of the architecture with desired RS knowledge is achieved using Low-Rank Adaptation (LoRA)~\cite{hu2021lora}, ensuring training efficiency while retaining essential knowledge for contextual understanding within the remote-sensing reasoning framework of GeoChat. This efficient training positions GeoChat as a prominent Vision LLM in remote sensing, particularly excelling in multimodal conversational capabilities with high-resolution RS imagery for grounding, region-specific, and image-level queries. A simple question-and-answer conversation with GeoChat is presented in Fig. \ref{GAI_Overview} for demonstration.

\textbf{TimeGPT for Time Series Data.} 
Time series data plays a pervasive role in the field of geoscience. Given the intricate dynamics and ever-evolving nature of Earth's systems, continuous monitoring and analysis of temporal changes become imperative. The modeling and forecasting of temporal dimensions are pivotal elements in comprehending and predicting trends across diverse geoscientific phenomena encompassing climatic patterns, seismic activity, ocean currents, urban morphology, and an array of other intricate variables. These geoscientific forecasts are crucial for risk assessment, climate analysis, resource management, and urban planning. It predicts natural hazards, aids in sustainable resource use, ensures public safety, supports energy production, and informs policymaking. To address the need for real-time accurate forecasts, Garza {\it et al.}\cite{garza2023timegpt} introduced TimeGPT, the first foundation model for time series forecasting. The pre-trained TimeGPT framework is capable of generating accurate zero-shot predictions on unseen datasets through efficient transfer learning. The foundation of the TimeGPT framework adopts a similar philosophy to the large language models. The architectural design incorporates the self-attention mechanism from the Transformers model~\cite{vaswani2017attention} for conducting zero-shot inference on novel time series. Within the encoder-decoder Transformer architecture of the TimeGPT framework, the input data, consisting of historical and optional exogenous data, is modeled. The encoder's attention mechanism learns diverse properties from the inputs, which are subsequently conveyed to the decoder for forecasting. The sequence of predictions concludes upon reaching the user-defined forecast horizon. Additionally, TimeGPT, coupled with conformal prediction techniques~\cite{vovk2005conformal, vovk2017nonparametric}, produces probabilistic forecasts and conducts anomaly detection without the need for specific dataset training, a significant advantage for small-sample size geoscience datasets. However, TimeGPT still do not work with spatio-temporal data that are omnipotent in geoscience domains.

\subsection{Physics-Informed Neural Networks in GeoScience}
Scientific machine learning (SciML) and Physics-informed neural networks (PINNs) are flourishing research areas of scientific computing aiming to build algorithms that combine purely data-driven methods with model-driven ones. Deep learning methods are good at discovering hidden patterns in large amounts of structured data for computer vision tasks and large language modeling tasks~\cite{goodfellow2016deep}. However, during the model training, very little is known about the `physics' or underlying (dynamical) systems or processes that one wants to learn from. This physical knowledge is usually in the form of an ordinary differential equation (ODE), a partial differential equation (PDE), or any other equation having an analytical or numerical solution. PINNs build a model to understand a physical process in an unsupervised manner by including the ODEs and PDEs within the loss function of the neural network using the training stage~\cite{raissi2019physics, rasht2022physics, yang2021b, jin2021nsfnets, mao2020physics}. Automatic differentiation (AD) is used to compute the gradient of the loss function and also to compute the derivatives of the output(s) of the network over the inputs. Once training is done, PINNs can quickly evaluate the solution of the chosen ODEs and PDEs at any point in the domain of interest. Thus, PINNs have been used in ML literature as a potential candidate for solving differential equations~\cite{lu2021deepxde} and prediction problems in climate science~\cite{elabid2022knowledge, dutta2023van}. A classical problem in geophysics is to solve the wave equation for a given number of different sources (viz. forcing terms) which has been solved using PINNs in the recent literature~\cite{moseley2020solving, pu2021solving}. To illustrate how PINNs work, we consider a simple PDE with two independent variables, here denoted with $t$, $x$ and $u(t)$ is the dependent variable. We are interested in $f$ which is a generic linear or nonlinear function of $u(t)$, and $\alpha$ is the free parameter:
\begin{equation*}
    \frac{\partial u(t,x)}{\partial t} + \frac{\partial u(t,x)}{\partial x} = f\left(u\left(t, x; \alpha\right)\right)
\end{equation*}
We write it more formally as:
\begin{align*}
    & \operatorname{PDE}\left(u\left(t, x\right)\right) = 0 \\
    &  \Rightarrow \frac{\partial u(t, x)}{\partial t} +\frac{\partial u(t, x)}{\partial x} - f\left(u\left(t, x; \alpha\right)\right) = 0 \\
    & \operatorname{IC:} \; u(t = 0 ) = u_0(x) \\
    & \operatorname{BC:} \; u(x = x_0 ) = u_{x_0}(t); \; u(x = x_1) = u_{x_1}(t)
\end{align*}
where we have specified the initial conditions (IC) and boundary conditions (BC). Then, a simple PINN model for the above PDE equation can be visualized in Fig. \ref{GAI_Models_Geosciece}, where $\mathcal{L}$ denoting the loss function and $f_{\theta}$ a simple feed-forward neural network for forward modeling. For a detailed discussion, readers can see~\cite{karniadakis2021physics}. An application of such a scenario in geoscience problems is the classical problem of traveltime tomography where the Eikonal equation is used as the PDE and PINNs have shown to represent an appealing solution~\cite{bin2021pinneik}. Knowledge-based deep learning is an example of a PINN-based structure used for modeling the El Niño dataset~\cite{elabid2022knowledge} whereas the Van der Pol neural network is used for multi-step ahead forecasting of seismic waves, temperature, and wind speed using PINN framework~\cite{dutta2023van}. Other emerging applications of PINNs in the field of geoscience are use cases of time-frequency domain wave equations~\cite{karimpouli2020physics}, Navier-Stokes equation~\cite{ranade2021discretizationnet, hu2024physics}, and inverse problems in geophysics~\cite{zhang2023seismic, depina2022application}.

\section{Benchmark Datasets and Resources in Geoscience GAI} \label{benchmarks}

Data types in geoscientific fields range from space and air to ocean and ground data, among many others. A non-exhaustive list of datasets, tools and resources for GAI in geoscience is given in Table \ref{tab:software-libraries}. We highlight in Table \ref{Geoscience_data_type} a refined perspective on the multifaceted nature of these datasets.

\begin{table*}[!t]
\caption{open-source datasets, tools, and resources for AI applications in Geoscience Data}
\label{tab:software-libraries}
\centering
\begin{tabular}{|l|p{70mm}|p{55mm}|r|}
\hline
\textbf{Datasets \& Toolbox} & \textbf{Brief Description}  & \textbf{Link} &\textbf{Ref.}\\
\hline
\textit{GeoImageNet}  & A benchmark dataset consisting of multisource
natural features (e.g., basins, bays, islands, lakes, ridges, and valleys) for supervised machine learning and AI in Geoscience. & \url{https://github.com/ASUcicilab/GeoImageNet} &~\cite{li2023geoimagenet}
\\ \hline
\textit{BigEarthNet}  & A benchmark archive consisting of over 590k
pairs of Sentinel-1 and Sentinel-2 image patches that were annotated with multi-labels of the CORINE Land Cover types to support deep learning studies in Earth remote sensing. & \url{https://bigearth.net/} &~\cite{sumbul2019bigearthnet} \\ \hline
\textit{EarthNets}  & An open-source platform that links to
hundreds of datasets, pre-trained deep learning models, and various tasks in Earth Observation. & \url{https://earthnets.github.io/} &~\cite{xiong2022earthnets}
\\ \hline
\textit{Microsoft Building Footprints}  & Microsoft Maps \& Geospatial teams released open building footprints datasets in
GeoJSON format in the United States, Canada, and Australia, as well as many countries in Africa and
South America. & \url{https://www.microsoft.com/en-us/maps} &~\cite{heris2020rasterized}
\\ \hline
\textit{ArcGIS Living Atlas}  & A large collection of geographic
information (including maps, apps, and GIS data layers) from around the globe.
It also includes a set of pre-trained deep learning models for geospatial applications such as land use classification, tree segmentation, and building footprint extraction. & \url{https://livingatlas.arcgis.com/} &~\cite{west2018gis}
\\ \hline
\textit{MoveBank}  & A publicly archived platform containing
over 300 datasets that describe the movement behavior of 11k animals. & \url{https://www.movebank.org/} &~\cite{kranstauber2011movebank}
\\ \hline
\textit{Geolife GPS Trajectories}  & An open
dataset contains 17,621 GPS trajectories by 182 users in a period of over three years with activity labels such as shopping, sightseeing, dining, hiking, and cycling. & \url{https://www.microsoft.com/en-us/download/details.aspx?id=52367} &~\cite{zheng2010geolife}
\\ \hline
\textit{Travel Flows}  & A multiscale dynamic
origin-to-destination population flow dataset (aggregated at three geographic scales: census tract, county, and state; updated daily and weekly) in the U.S. during the COVID-19 pandemic. & \url{https://github.com/GeoDS/COVID19USFlows} &~\cite{kang2020multiscale}
\\ \hline
\textit{Google Earth Engine}  & A multi-petabyte catalog of satellite imagery and geospatial datasets with planetary-scale analysis capabilities
and the Earth Engine API for geo-computation and analysis is available in JavaScript and Python, e.g., the geemap package & \url{https://earthengine.google.com/} &~\cite{wu2020geemap}
\\ \hline
\textit{ArcGIS GeoAI Toolbox}  & A ready-to-use tool for training and using machine/deep learning models that perform classification and regression on geospatial feature layers, imagery, tabular, and text
datasets. & \url{https://pro.arcgis.com/en/pro-app/latest/tool-reference/geoai} &~\cite{fleming2022toxpi}
\\ \hline
\textit{CyberGISX}  & An open platform for developing and sharing open educational resources (e.g., Jupyter Notebooks) on computationally intensive and reproducible geospatial analytics and workflows powered by CyberGIS middleware and cyber infrastructure. & \url{https://cybergisxhub.cigi.illinois.edu/} &~\cite{wang2013cybergis}
\\ \hline
\end{tabular}
\end{table*}

\begin{figure*}[htb] 
\centering
\includegraphics[width=0.98\textwidth]{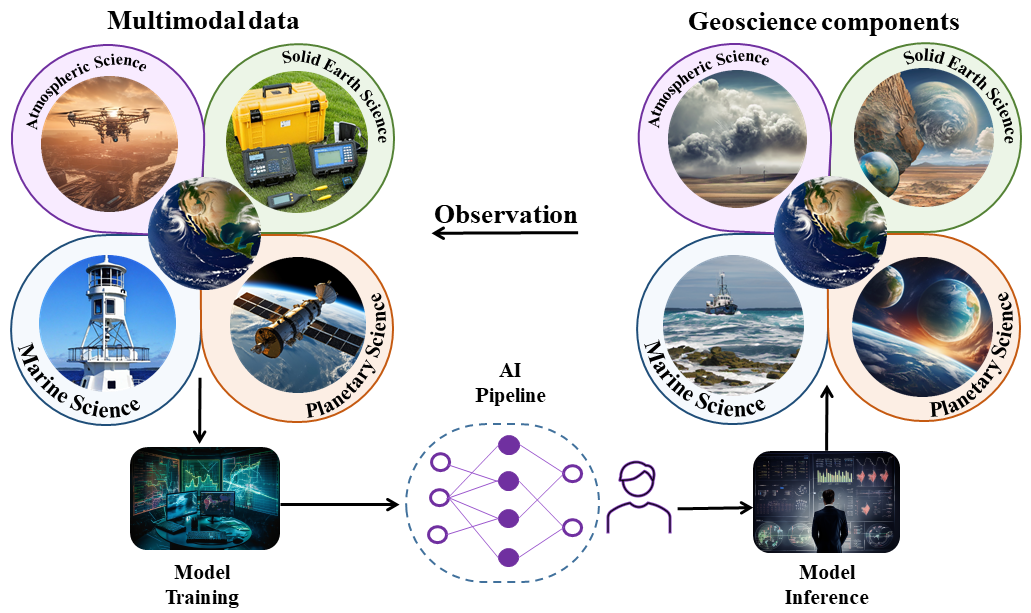}
\caption{Connectionism between Geoscience foundation models and their broad usage in geosciences and related technologies. (Left) Data collected from various sources in geoscience, including space, air, ground, and ocean machinery, provide multimodal data for model training that analyzes the data using GAI models and performs tasks in real-time that are specified by the user to support geoscience research and advancements.}
\label{Fig_challenges}
\end{figure*}

\begin{table*}[!t]
\caption{Geoscience data types, their descriptions, and sources.}
\label{Geoscience_data_type}
\centering
\begin{tabular}{|p{25mm}|p{60mm}|p{40mm}|p{35mm}|}
\hline
\textbf{Data type} & \textbf{Data Description}  & \textbf{Geoscience Components} &\textbf{GAI Tools}\\
\hline
Climate and Weather data~\cite{de2020open}  & This type of data comprises data on temperature, precipitation, solar radiation, wind, air, pressure, etc. These datasets are used for weather forecasting and understanding climate change  & Atmospheric Science (air-based data) and Marine Science (ocean-based data) & 
 TimeGPT, PINNs \\ \hline

Time series and geospatial data~\cite{lewis2017australian} & Climate and seismic datasets are mostly time series data incorporating temporal observations. Geospatial data are facilitated through Geographic Information Systems (GIS), e.g., geographic coordinates, spatial databases, etc. & Planetary Science (space-based data) & TimeGPT, PINNs, MetroGAN \\ \hline

Satellite and remote sensing data~\cite{navalgund2007remote} & Satellite imagery and remote sensing data encompass optical, infrared, radar, and lidar datasets. These are used to understand land cover land use (LULC) classification, vegetation, and environmental changes. & Planetary Science (air-based data) & TimeGPT, PINNs, CityGAN, RidgeGAN \\ \hline

Geological and geophysical data~\cite{bressan2020evaluation} & These datasets include seismic readings, gravity measurements, sea-surface temperature, well logs, and geological maps. It can be used for understanding subsurface structures and rock properties. & Solid Earth Science and Marine Science & Bayesian GAN, U-Net, SeismoGen \\ \hline

Environmental data~\cite{bressan2020evaluation} & These datasets include air and water quality, air pollution, and environmental hazards data. It can be used for monitoring environmental conditions and pollution control. & Atmospheric Science and Solid Earth Sciences & PINNs, Graph Neural Networks \\ \hline

Text and literature data~\cite{zhang2023geodeepshovel} & Scientific literature on geoscience, research publications, field reports, etc., serve as valuable resources for the geoscience community. These datasets are useful for training LLMs for providing domain knowledge & Geoscience books and publications & K2, GeoChat \\ \hline
\end{tabular}
\end{table*}

As depicted in Fig. \ref{Fig_challenges}, data from geoscience components brings multimodal observations that are used for model training and inference. Generative AI models play a vital role in analyzing data from Atmospheric Science, Planetary Science, Solid Earth Science, and Marine Science~\cite{zhang2023geoscience}. Space-based data (Planetary Science) gives an overview of the Earth's surface and atmosphere. Air-based data (Atmospheric Science) provide insights about space and terrestrial observations using high-quality data acquisition, Ground-based data (Solid Earth Science) is collected from geographical or geological occurrences. Lastly, ocean-based data (Marine Science) gives insights into the evaluation of marine and climate conditions. Geoscience foundation models are transforming the field of research with profoundly significant frontiers and their applications are evident in Earth system~\cite{artificialhandbook}. The availability of geoscience data will lead to some excellent GAI models, especially with open-source efforts from both industries and academia.

\section{Challenges of Using Generative AI in Geoscience}

\begin{figure*}[htb] 
\centering
\includegraphics[width=0.98\textwidth]{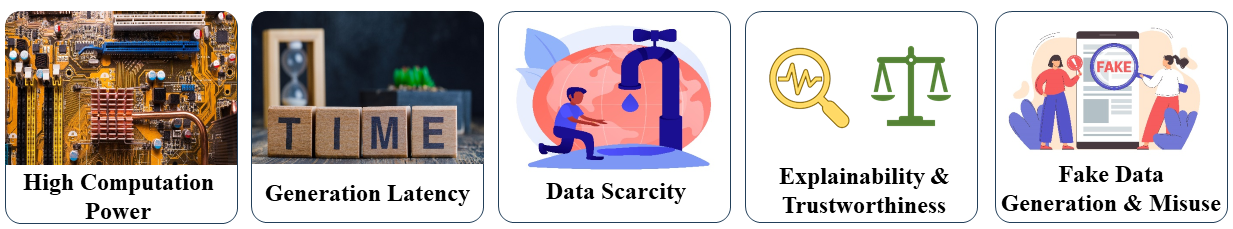}
\caption{Some challenges of using generative artificial intelligence in geoscience. GAI models use high computational power, consume time during training, lack transparency, and sometimes are misused for fake data generation. These are the potential threats for adapting GAI models for geoscientific advancements.}
\label{challenges}
\end{figure*}

The use of traditional machine learning algorithms in geoscience applications is quite limited due to a variety of factors. First of all, the nature of geoscience processes poses certain inherent obstacles. For example, geoscience objects generally have amorphous boundaries in time and in space, while objects in other domains are more crisply defined~\cite{chakraborty2023ten}. Hence, the nature of geoscience objects that exist in continuous spatiotemporal fields is much more complex than that found in discrete spaces that machine learning algorithms typically deal with. Moreover, geoscience phenomena are generally multi-variate, following non-linear relationships (e.g., chaotic), having non-stationary properties, and containing unusual but fascinating events. In addition to the intrinsic difficulties of geoscience processes, the methodologies for gathering geoscience data present additional difficulties for machine learning algorithms. This includes the presence of samples at multiple resolutions, with varying degrees of incompleteness, and uncertainties. While generative AI has shown immense potential in various domains, its resource-intensive nature may hinder real-time use and scalability. In fact, large-scale generative AI models usually require a significant amount of computational resources and electrical power to operate, resulting in elevated energy consumption and significant carbon emissions. This may, unfortunately, restrict their usage in many real-world applications. Future efforts are then needed to design novel and efficient AI architectures capable of generating high-quality data points in real time, which is needed for constrained platforms. GAI tools use high computation power that has huge environmental impact and cost inefficient. ChatGPT-based models are ``black-box'' in nature which opens doors for future researchers and potential applications in geoscience. Domain knowledge in geoscience plays a vital role in building GAI models. However, inadequate data for any scientific class may include biases in GAI models and this could be a potential risk for geoscientists during decision making. As generative artificial intelligence models can be considered to be still in their infancy, here are some of the key challenges that still have to be tackled to ensure applicability in geoscience (also see Fig.~\ref{challenges}):
\begin{itemize}
\item {\bf High Computational Power:} Generative AI models usually consist of billions of parameters. Large-scale computing infrastructures are usually necessary to develop such large models. For example, diffusion models require millions or billions of images to train. To train with such large datasets, massive computing power (clusters with hundreds of TPUs/GPUs) may be needed.
\item {\bf Generation Latency:}  There is often some latency present in the time it takes to generate a new sample, mainly due to the large scale of generative AI models. If we take as an example the diffusion models~\cite{yang2023diffusion}, which are popular in the generation of high-quality samples, their sampling speed is, however, known to be slow.
\item {\bf Data Scarcity:} Among the main challenges in using AI in general is the lack of large datasets for training. Generative AI models usually require a large amount of high-quality and unbiased data to operate. Lots of research has indeed been devoted to coping with the problem of data scarcity in AI. Although generative AI models can be used to produce synthetic data for training, other strategies can also be devised for scenarios with limited data. This includes few-shot learning~\cite{song2023comprehensive}, transfer learning~\cite{zhuang2020comprehensive}, and domain adaptation~\cite{yu2023comprehensive}, which offer the potential of enhancing the AI performance when the data is scarce.
\item {\bf Explainability and Trustworthiness:}
Lack of transparency and explainability are among the other barriers to the ubiquitous integration of generative artificial intelligence in many real-world problems. A possible step towards trustworthy AI is to develop explainable AI. Explainable AI refers to models capable of generating decisions that a human could understand and interpret. This would strengthen the trust. TrustLLM can be a potential solution for this problem for the scientific community and geoscientists~\cite{lichao2024TrustLLM}. This is not only limited to privacy-sensitive research problems like in medical science but also to economically critical operational decisions.
\item {\bf Fake Data Generation and Misuse:}
These technologies can also be used in the context of geoscience and remote sensing for wicked cases to create fake satellite images, among many others~\cite{patel2023generative}. For example, fake data generation in geoscience may result for a non-expert user of information and evidence of something that seldom exists and hides the truth. A real-life example could be a country that is trying to hide some land surface changes in an area to mask human right violations~\cite{rodrigues2020legal}. AI-generated data could replace the same places as real satellite imagery and can be possibly misused to fake the reality of the situation from the public.   
\end{itemize}
\section{Discussion} \label{discussion} 
The integration of generative AI promises a paradigm shift in geoscience. As we strive for enhanced safety, efficiency, and sustainability, generative AI emerges as a key differentiator. It can help build a sustainable, prosperous, and technologically advanced research area in geoscience. It can accelerate the transition towards a greener, more resilient landscape and environmental sustainability. However, with every innovation comes skepticism. The development of generative models raises serious concerns about the potential emergence of super-intelligent machines without adequate safeguards. Generative AI can indeed unintentionally produce incorrect information due to biases in the training data. In this context, the European Union has just released some future regulations on the use of Artificial Intelligence and Large Language Models. The provisional regulations indicate that foundation models must comply with specific transparency obligations. 
Moreover, generative models usually require a large amount of high-quality data to operate. Other issues concern the latency for generating high-quality samples and the massive computing power that is needed to train generative models. Lack of explainability is also among the barriers to the ubiquitous integration of generative artificial intelligence. So, in the end, can we really trust the outcomes of generative AI models for critical operational decisions in geoscience?  This calls for more responsible, explainable, and trustworthy AI models in different fields, including geoscience.

When using generative artificial intelligence, it is then important to be aware of the risks and the challenges the models are facing. For instance, to mitigate the challenge of the lack of data in geoscience, one can impose some specific constraints to ``guide'' the model to learn faster and better during the training. This is an effective way not only to mitigate the shortage of training data but also to increase models’ generalizability and to ensure the physical plausibility of the results. This is basically the idea behind PINNs that take into consideration prior knowledge about the physical laws that govern the data in the learning process. PINNs overcome the low data availability issue~\cite{elabid2022knowledge} in which GANs and Transformers may lack robustness~\cite{sasal2022w}, rendering them less effective. PINNs are shown to be effective and popular in geophysics~\cite{hao2022physics}.
 
\section{Conclusion} \label{conclusion} 
In this paper, we discussed the potential of generative AI and large language models for understanding and predicting geoscience and earth system dynamics. We introduced generative AI as a remarkable tool with amazing capabilities across a broad spectrum of tasks. Then, we discussed the integration of generative AI in geoscience, arguing the relatively unexplored territory with great potential. We also enumerated a non-exhaustive list of potential applications of generative AI in geoscience. As every innovation comes with skepticism, we then presented some of the key challenges that still have to be tackled to ensure the applicability of generative AI in geoscience. To further explain the use of generative AI in geoscience, a concrete example was given, describing models based on GANs for data generation in geoscience. The model is a seismic waveform synthesis technique that exploits GAN for seismic data generation. Moreover, several examples of large language models capable of understanding geoscience and generating new geoscience ideas are also described. Lately, we have discussed the importance of mitigating the AI challenges and, specifically, the lack of data in geoscience to train the AI models by imposing specific constraints to ``guide'' the model training. This is an effective way to mitigate the shortage of training data, to increase models’ generalizability and to ensure the physical plausibility of the results (as done in PINNs). Other limitations of GAI models in geoscience involve scalability, interpretability, trustworthiness, social biases, and fake data generation have also been highlighted. Improved scalability and generalizability can make GAI models tackle the challenges of Earth's systems.

The use of generative AI in geoscience holds the promise of transforming various aspects of the industry, from reservoir analysis and production optimization to safety protocols and collaborative decision-making. As AI technology continues to advance, the integration of generative AI into geoscience could then mark a significant step forward in the industry’s pursuit of innovation and optimization. Although generative AI can offer elegant solutions to many technical challenges in geoscience, it is important to note that generative AI is not a complete solution that can replace all human experience and expertise. Some technical problems may necessitate a thorough comprehension of the underlying phenomena best mastered by human experts. 

In conclusion, AI is promised to play a pivotal role in the future of geoscience. It is a powerful tool that can transform the way we interpret and understand geophysical data. While there are challenges to overcome, the potential benefits far outweigh the hurdles. As we continue to harness the power of generative AI, we can look forward to a new era of innovation and progress in geoscience. Going forward, joint efforts between geoscientists, computer scientists, and the AI community can make discoveries in pressing challenges like climate change, environmental hazards, and sustainability. This paper is intended to provide guidance but does not claim to be comprehensive or conclusive in any way. 
\vspace{-0.1cm}

\section*{Acknowledgments}
The support of TotalEnergies is fully acknowledged. Abdenour Hadid (Professor, Industry Chair at SCAI Center of Abu Dhabi) is funded by TotalEnergies collaboration agreement with Sorbonne University Abu Dhabi. The authors also acknowledge open-source generative AI tools that have been used for the generations of images used in this paper. T.C. acknowledges Madhurima Panja of IIIT Bangalore, India for valuable insights and suggestions. 

\bibliographystyle{IEEEtran}
\bibliography{refs}

\end{document}